\documentclass{elsarticle}
\usepackage{times}
\usepackage{natbib}
\usepackage{hyperref}  
\usepackage[lmargin=.95in,rmargin=.95in,tmargin=0.9in,bmargin=0.7in, showframe=false]{geometry}
\usepackage[T1]{fontenc}
\usepackage{amsfonts}
\usepackage{amsmath}
\usepackage{color}

 \usepackage{array,multirow,graphicx}
\usepackage{multicol}
\usepackage{picture}
\usepackage{amsmath}
\usepackage{chngpage}
\usepackage{fixltx2e}
\usepackage[utf8]{inputenc}
\usepackage[english]{babel}

\newtheorem{theorem}{Theorem}

\newtheorem{definition}[theorem]{Definition}
\newtheorem{lemma}{Lemma}

\usepackage{amsmath}
\makeatletter
\newcommand{\xMapsto}[2][]{\ext@arrow 0599{\Mapstofill@}{#1}{#2}}
\def\Mapstofill@{\arrowfill@{\Mapstochar\Relbar}\Relbar\Rightarrow}
\makeatother
\DeclareMathOperator*{\esssup}{ess\,sup}
\usepackage{hyperref}
\newcommand*{\QEDB}{\hfill\ensuremath{\square}}%

\begin{document}

\begin{frontmatter}

\date{}

\title{A Novel Approach to Quantification of Model Risk for Practitioners}

\author[rvt]{Z.~Kraj\v{c}ovi\v{c}ov\'{a}*}
\ead{zkrajcov@udc.es}

\author[els]{P. P.~ P\'{e}rez Velasco}

\author[focal]{C.~V\'{a}zquez}

\cortext[cor1]{Corresponding Author}

\address[rvt]{Departmento of Mathematics, University of A Coruña, Spain}
\address[els]{Model Risk Division, Banco Santander, Spain}
\address[focal]{Departmento of Mathematics, University of A Coruña, Spain}

\begin{abstract}
Models continue to increase their already broad use across industry as well as their sophistication. Worldwide regulation oblige financial institutions to manage and address model risk with the same severity as any other type of risk, e.g. \cite{reserve2011supervisory}, which besides defines model risk as the potential for adverse consequences from decisions based on incorrect and misused model outputs and reports. Model risk quantification is essential not only in meeting these requirements but for institution's basic internal operative. It is however a complex task as any comprehensive quantification methodology should at least consider the data used for building the model, its mathematical foundations, the IT infrastructure, overall performance and (most importantly) usage. Besides, the current amount of models and different mathematical modelling techniques is overwhelming.\\
 
Our proposal is to define quantification of model risk as a calculation of the norm of some appropriate function that belongs to a Banach space, defined over a weighted Riemannian manifold endowed with the Fisher--Rao metric. The aim of the present contribution is twofold: Introduce a sufficiently general and sound mathematical framework to cover the aforementioned points and illustrate how a practitioner may identify the relevant abstract concepts and put them to work.
\end{abstract}


\end{frontmatter}

\section{Introduction}

Models are simplifying mappings of reality to serve a specific purpose aimed at applying mathematical, financial and economic theories to the available data. They deliberately focus on specific aspects of the reality and degrade or ignore the rest. Understanding the capabilities and limitations of the underlying assumptions is key when dealing with a model and its outputs. According to the \cite{reserve2011supervisory} model risk is defined as\\

\begin{adjustwidth}{1cm}{1cm}
{\it "[\dots] the potential for adverse consequences from decisions based on incorrect or misused model outputs and reports. Model risk can lead to financial loss, poor business and strategic decision making, or damage to bank's reputation"\\ }
\end{adjustwidth}

Fed then identifies the two main reasons for model risk (inappropriate usage and fundamental errors). Further, they state that model risk should be managed and addressed with the same severity as any other type of risk and that banks should identify the sources of model risk and assess their magnitude. Fed also emphasizes that expert modelling, robust model validation and a properly justified approach are necessary elements in model risk moderation, though they are not sufficient and should not be used as an excuse for not improving models.  
\\

In spite of the rise of awareness of model risk and understanding its significant impact, there are no globally defined industry and market standards on its exact definition, management and quantification, even though a proper model risk management is required by regulators. \\

Within the finance literature, some authors have defined model risk as the uncertainty about the risk factor distribution (\cite{gibson2000model}), the misspecified underlying model ( \cite{cont2006model}), the deviation of a model from a 'true' dynamic process (\cite{branger2004model}), the discrepancy relative to a benchmark model ( \cite{hull2002methodology}), and the inaccuracy in risk forecasting that arises from the estimation error and the use of an incorrect model (\cite{boucher2014risk}).  Model risk has been classified previously in all asset classes, see \cite{morini2011understanding} for interest rate products and credit products, \cite{christodoulakis2008validity} for portfolio applications, \cite{saltelli2013sensitivity} for asset backed securities, and \cite{boucher2014risk} for relation to measuring marker risk.
\\

The quantification, as an essential part of model risk management, is required for a consistent management and effective communication of model weaknesses and limitations to decision makers and users and to assess model risk in the context of the overall position of the organization.
The quantification of model risk should consider the uncertainty stemming from the selection of the mathematical techniques (e.g. focusing on fitting a normal distribution hence leaving aside other distribution families), the calibration methodology (e.g. different optimization algorithms may derive different parameter values), and from the limitations on the sample data (e.g. sparse or incomplete database). \\

Model risk quantification poses many challenges that come from the high diversity of models, the wide range of techniques, the different use of models, among others. Some model outputs drive decisions; other model outputs provide one source of management information, some outputs are further used as an inputs in other models. Additionally, the model outputs may be completely overridden by expert judgement. Not to mention that in order to quantify model risk you need another model, which is again prone to model risk.\\

The most relevant areas of analysis for the quantification of model risk are: data and calibration, model foundations, model performance, IT infrastructure, model use, controls and governance, and model sensitivity. 
The model may be fundamentally wrong due to the errors in theoretical foundation and conceptual design that emerge from incorrect logic or assumptions, model misspecification or omission of variables. Data quality issues, inadequate sample sizes and outdated data contribute to model performance issues such as instability, inaccuracy or bias in model forecasts. Model risk also arises from inadequate controls over the model use. Flawed test procedures or failure to perform consistent and comprehensive user acceptance tests can lead to material model risk. To name just a few.\\ 

The focus of this paper is on developing a novel approach for quantifying model risk within the framework of differential geometry (\cite{murray1993differential})  and information theory (\cite{amari1987differential}). In this work we introduce a measure of model risk on a statistical manifold where models are represented by a probability distribution function. Differences between models are determined by the geodesic distance under the Fisher--Rao metric. This metric allows us to utilize the intrinsic structure of the manifold of densities and to respect the geometry of the space we are working on, i.e. it accounts for the non--linearities of the underlying space. \\ 

The rest of this paper is structured as follows. In section 2, we summarize basic facts about Riemannian geometry and introduce the terminology used throughout the paper. Modeling process steps and a general description of our proposed method for quantification of model risk are presented in Section 3, which is followed by a detailed discussion on the main quantification steps. Section 4 to 6 describe the construction of the neighbourhood containing material variations of the model, and the definition and construction of the weight function. The model risk measure is then defined and explained in Section 7. Section 8 provides some final conclusions and directions for future work, and finally, the Appendix contains the proofs of the main results.  


\section{Background on Riemannian Geometry}\label{sec:back}

This section introduces the necessary notation for the rest of the paper. The details can be found among other standard references in \cite{amari1987differential} or \cite{murray1993differential}.\\

$\mathcal{M}$ is a compact and connected manifold without boundary equipped with a Riemannian metric $<\cdot,\cdot>$ and a Riemannian connection $\bigtriangledown$, with $T_{p}\mathcal{M}$ the tangent space at $p\in \mathcal{M}$. The distance $d$ between $p,q \in \mathcal{M}$ is given by
\begin{eqnarray*}
d(p,q):= \inf_{\gamma} \int_{a}^{b} ||\gamma'(t)||dt,
\end{eqnarray*}
where $\gamma$ ranges over all differentiable paths $\omega: [a,b]\rightarrow \mathcal{M}$ satisfying $\gamma(a)=p$ and $\gamma(b)= q$, and $||\gamma'||^2=\big<\gamma', \gamma'  \big>$. $(\mathcal{M}, d)$ is a metric space. \\

The Riemannian metric associates to each point $p\in \mathcal{M}$ an inner product $<\cdot,\cdot>_p$ on $T_p\mathcal{M}$. 
One natural metric on the Riemannian manifold $\mathcal{M}$ is the Fisher--Rao information metric (\cite{rao1945information})
\begin{eqnarray}\label{eq:Fisher}
  I_{ij}(p)=g_{ij}(p)= \mathbb{E} \Big[ \dfrac{\partial \log (p)}{\partial x^{i}} \dfrac{\partial \log (p)}{\partial x^{j}} \Big]=\int p \dfrac{\partial \log (p)}{\partial x^{i}}\dfrac{\partial \log (p)}{\partial x^{j}}dx
\end{eqnarray}
The $det\, I(p)$ represents the amount of information a sample point conveys with respect to the problem of estimating the parameter $\mathbf{x}$, and so $I(p)$ can be used to determine the dissimilarities between distributions. \\

Under a square--root representation, the Fisher--Rao metric becomes the standard $\mathbb{L}^2$ metric and the space of probability density functions becomes the positive orthant of the unit hypersphere in $\mathbb{L}^2$ (\cite{lang2012fundamentals}).
The square--root mapping is defined as a continuous mapping $\phi: \mathcal{M}\rightarrow \Psi$ where $\Psi$ is the space containing the positive square--root of all possible density functions. Using this mapping, we define the square--root transform of probability density functions as $\phi(p)=\psi=\sqrt{p}$:
\begin{eqnarray*}
\Psi=\Big\{ \psi(x) \Big| \int_{\mathcal{X}}|\psi(s)|^2ds=1, \forall s \,\, \psi(s)\geq 0, x\in \mathcal{X} \Big\},
\end{eqnarray*}
where $\mathcal{X}$ is the sample space. In this case, the associated natural Hilbert space, $\mathcal{H}$, equipped with a symmetric inner product, $g_{ij}$, induces a spherical geometry, i.e. the sum $\sum (\sqrt{p})^2$ is equal to unity (\cite{lang2012fundamentals}). If the density function is parametrized by the set of parameters $\mathbb{\theta}=(\theta^1,\dots, \theta^n)$ then for each value of $\theta^i$ we have a corresponding point on the unit sphere $S$ in $\mathcal{H}$. In this setting the geodesics are available in closed form and can hence be computed quickly and exactly.
For any two tangent vectors $v_1, v_2\in T_{\psi}\Psi$, the Fisher--Rao metric is given by
\begin{eqnarray}\label{eq:FR}
\big<v_1,v_2\big>&=& \int_{\mathbb{R}} v_1(s)v_{2}(s)ds =  \Bigg< \dfrac{\partial \{ p(\cdot, \theta) \}^{1/2}}{\partial \theta_i}, \dfrac{\partial \{ p(\cdot, \theta) \}^{1/2}}{\partial \theta_j} \Bigg> =\dfrac{1}{4}g_{ij}
\end{eqnarray}
The geodesic in the direction $v$ on the sphere and the distance given two points $\psi_1, \psi_2$ belonging to the sphere are given by
\begin{eqnarray*}
\gamma(t)&=&\cos(t||v||)\psi+\sin(t||v||)\dfrac{v}{||v||}\\
d(\psi_1, \psi_2)&=&\cos^{-1}(\big< \psi_1, \psi_2 \big>)
\end{eqnarray*}

Since the compactness of $\mathcal{M}$ implies geodesic completeness (\cite{chavel2006riemannian}), there exists for every $p\in \mathcal{M}$ and $v\in T_{p}\mathcal{M}$ an unique geodesic $\gamma: \mathbb{R}\rightarrow \mathcal{M}$ satisfying $\gamma(0)=p$ and $\gamma'(0)=v$. Moreover, the Hopf--Rinow Theorem (\cite{HopfRinow}) ensures that any two points $p,q\in \mathcal{M}$ can be joined by a minimal geodesic of length equal to the distance between the points, $d(p,q)$. Through the geodesic $\gamma$, one can define the exponential map $\exp_p: T_{p}\mathcal{M}\rightarrow \mathcal{M}$ by 
\begin{eqnarray*}
\exp_{p} tv:= \gamma(t), \,\,\,\, \forall t\in \mathbb{R}, \, \forall v\in \mathcal{M}.
\end{eqnarray*}
The exponential map for square--root transformation (\cite{joshi2007riemannian}) has the form
\begin{eqnarray*}
\exp_{\psi_i} tv:= \cos\Big(||tv||_{\psi_i}\Big)\psi_i + \sin\Big(|| tv||_{\psi_i}\Big)\dfrac{tv}{||tv||_{\psi_i}},
\end{eqnarray*}
where $v\in T_{\psi_i}(\Psi)$. The inverse exponential map from $\psi_i$ to $\psi_j$ is given by
\begin{eqnarray}\label{eq:exponential}
\exp_{\psi_i}^{-1} (\psi_j):= \dfrac{d(\psi_1, \psi_2)}{\sin(d(\psi_1, \psi_2))}\Big(  \psi_j - \cos(d(\psi_1, \psi_2))\psi_i \Big).
\end{eqnarray}

An open set $U\subset \mathcal{M}$ is said to be a normal neighbourhood of $p_0 \in U$, if $\exp_{p_0}$ is a diffeomorphism on a neighbourhood $V$ of the origin of $T_{p_0}\mathcal{M}$ onto $U$, with $V$ such that $tv\in V$ for $0\leq t\leq 1$, if $v\in V$.


\section{Modeling Process Steps and Quantification of Model Risk}

There are different types and aspects of model risk that tend to easily overlap, co--occur, or co--vary. 
In this context, we propose four rough model creation steps: Data, Calibration, Model Selection and Testing, and Implementation and Usage. This may occur in an iterative fashion, but they result in a general linear flow that ends with institutional use (implementation and maintenance) to direct decision making (often encoded into an IT system). Limitations in any of these areas can impair reliance on model results.

\begin{enumerate}
\item {\bf Data} refers to the definition of the purpose for modeling, the specification of the modeling scope, human and financial resources, the specification of data and other prior knowledge, their interpretation and preparation. The data may be obtained from both internal and external sources, and they are further prepared by cleaning and reshaping. Model risk may arise from data deficiencies in terms of both quality and availability, including, among others, error in data definition,  insufficient sample,
inaccurate proxies, sensitivity to expert judgments, or misinterpretation.

\item {\bf Calibration} includes the selection of the types of variables and the nature of their treatment, the tuning of free parameters, and links between system components and processes.
Estimation uncertainty may occur due to simplifications, approximations, flawed assumptions, inappropriate calibration, wrong selection of subset, errors in statistical estimation or in market benchmarks, computational or algorithmic limitations, or use of unobservable parameters.

\item {\bf Model Selection and Testing} involves the choice of the estimation performance criteria and techniques, the identification of model structure and parameters, which is generally an iterative process with the underlying aim to balance sensitivity to system variables against complexity of representation. Further, it is related to the conditional verification which includes testing the sensitivity to changes in the data and to possible deviations from the initial assumptions. In this step, model risk stems from, e.g., inadequate and incorrect modeling assumptions, outdated model due to parameter decalibration, model instability, 
or model misspecification. 

\item {\bf Implementation and Usage} refers to the deployment of the model into production which is followed by a regular maintenance and monitoring. Sources of model risk in this step include using the model for unintended purposes, luck of recalibration, IT failures, luck of communication between modelers and users, luck of understanding on model limitations.
\end{enumerate}

Quantification of model risk, from a best practice perspective, should be quick and reliable, without refitting or building models, without reference to particular structure and methodologies, and with prioritizing analysis (getting immediate assurance on shifts that are immaterial). Differential geometry and information theory offer a base for such an approach. In this framework, a model is represented by a particular probability distribution, $p: \mathcal{X}\rightarrow \mathbb{R}_{+}$ that belongs to the set of probability measures $\mathcal{M}$, so called statistical manifold, available for modelling. The manifold $\mathcal{M}$ can be further equipped with the information--theoretic geometric structure that, among other things, allows us to quantify variations and dissimilarities between probability distribution functions (models). \\

The set of possible probability measures may be further parametrized in a canonical way by a parameter space $\Theta$, $\mathcal{M}=\{p(x;\theta): \theta\in \Theta \}$. This set forms a smooth Riemannian manifold $\mathcal{M}$. Every distribution is a point in this space, and the collection of points created by varying the parameters of the model, $p\in \mathcal{M}$, gives rise to a hypersurface (a parametric family of distributions) in which similar distributions are mapped to nearby points. 
The natural Riemannian metric is shown to be the Fisher--Rao metric (\cite{rao1945information}) which is the unique intrinsic metric on the statistical manifold. It is the only metric that is invariant under re--parametrization, \cite{amari1987differential}.
\\

Let us consider a given model $p_0$ which can be uniquely parametrized 
using the vector $\theta_0=(\theta^1_0, \dots , \theta^n_0)$ over the sample space $\mathcal{X}$ and which can be described by the probability distribution $p_0=p(x;\theta_0)$. 
This probability distribution belongs to a set (family) of distributions $\mathcal{M}=\{p(x;\theta): \theta\in \Theta \subset \mathbb{R}^n\}$ that forms a model manifold. We assume that for each $x\in \mathcal{X}$ the function $\theta \mapsto p(x; \theta)$  is $C^{\infty}$. Thus, $\mathcal{M}$ forms a differentiable manifold and we can identify models in the family with points on this manifold. Choosing a particular model is the same as fixing a parameter setting $\theta\in \Theta$. 

\subsection*{Example}
To help fix ideas, we introduce an illustrative simple example and develop it further throughout the paper. Let $\mathbb{X}$ denote a vector of profit and loss, $P\& L$, over a two year time horizon (520 days) that is used to calculate the Value at Risk (VaR). VaR is derived from a distribution of $P\& L$ as the quantile loss at the portfolio level and is defined by
\begin{eqnarray*}
\mathbb{P}(\mathbb{X} \leq VaR)=1-\beta,
\end{eqnarray*}
where $\beta$ is the confidence level set to $99.9\%$. 
Assume that the given model considers $\mathbb{X}$ to be normally distributed $p_{0}=\mathcal{N}(\mu_0, \sigma_0)$ with parameters $\mu_0=2, \sigma_0=10$ once calibrated. This model belongs to a family of normal distributions that forms a differentiable manifold $\mathcal{M}=\{ p(x; \mu, \sigma): \mu \in \mathbb{R}, \sigma>0\}$ where $\mu$ is the mean and $\sigma$ is the standard deviation. Every point $p\in \mathcal{M}$ corresponds to a normal distribution $p(x, \theta)$ with $\theta=(\mu, \sigma)$.\\
 
In our univariate normally distributed case parametrized by a 2--dimensional space, $\theta=(\mu, \sigma)$, the Riemannian matrix defined by $\ref{eq:FR}$ is given by
\begin{eqnarray*}
I=[I_{ij}(\mu, \sigma)]=\begin{bmatrix}
\dfrac{1}{\sigma^2} & 0 \\
0 & \dfrac{2}{\sigma^2}
\end{bmatrix} = \begin{bmatrix}
0.01 & 0 \\
0 & 0.02
\end{bmatrix}.
\end{eqnarray*} 

\QEDB
\\

We define the model risk for a given model $p_0$ at the scale of an open neighbourhood around $p_0$ that contains alternative models that are not too far in a sense quantified by the relevance to (missing) properties and limitations of the model. 
The model risk is then measured with respect to all models inside this neighbourhood as a norm of an appropriate function of the output differences over a weighted Riemannian manifold endowed with the Fisher--Rao metric and the Levi--Civita connection\footnote{The Levi--Civita connection parallely transports tangent vectors defined at one point to another and is compatible with the geometry induced by the Riemannian metric (\cite{amari1987differential}). Additionally, for this choice of connection, the shortest paths are geodesics.}. The analysis consists of five steps: 
\begin{enumerate}
\item Embedding the model manifold into one that considers missing properties\footnote{Or properties not appropriately modelled, for which there is no consensus, cannot be adequately calibrated, among many others.} in the given model $p_0$.
\item Choosing a proper neighbourhood around the given model.
\item Choosing an appropriate weight function, that assigns relative relevance to the different models inside the neighbourhood.
\item Calculating the measure of model risk with respect to all models inside the neighbourhood, through the corresponding norm.
\item Interpretation of the measure with respect to the specific use of the model risk quantification.
\end{enumerate}
Each step addresses and aligns different limitations of the model and the uncertainty in various areas related to the model\footnote{Such as data, calibration, model selection, model performance, model sensitivity and scenario analysis, and most importantly the usage of the model}. In the following sections we further develop these steps and describe the intuition behind.  

\section{Neighbourhood Around the Model}

Recall that the given model $p_0$ belongs to a $n$--dimensional manifold $\mathcal{M}$ where each dimension represents different pieces of information inherited in $p_0$. To consider missing properties, the uncertainty surrounding the data and the calibration, the additional information about the limitations of the model, or wrong underlying assumptions, we may need to adjoin new dimensions to $\mathcal{M}$, and thus, consider a higher--dimensional space within which $\mathcal{M}$ is embedded. \\

The proper neighbourhood around $p_0$ we define with the help of the tangent space $T_{p_0}\mathcal{M}$ at a point $p_0$. $T_{p_0}\mathcal{M}$ is a vector space that describes a first order approximation, infinitesimal displacements or deformations on the manifold in the position of the point $p_{0}$. 

From a practical point of view, not all perturbations are relevant, thus taking into account the materiality with respect to the intended purpose of the model, its usage, business and market, we consider only a small subset of the tangent bundle.\\

Let $\mathcal{U}$ be the open set around $p_0$ of some normal neighbourhood $V$ such that
\begin{eqnarray*}
\mathcal{U} :=\{ tv\in V\subset T_{p_0}\mathcal{M}: 0< t \leq \alpha(v), \,\, v\in \mathcal{S}(p_0, 1) \text{ and normal coordinates are defined}\},
\end{eqnarray*}
where $\mathcal{S}(p_0, 1)=\{v\in T_{p_0}\mathcal{M}, ||v||=1 \}$ is a unit sphere on $T_{p_0}\mathcal{M}$.\\

The neighbourhood $\mathcal{U}$ includes the directions of all relevant perturbations of the model $p_0$ up to a certain level $\alpha(v)$. The level $\alpha(v)$ depends on the tangent vectors, since the degree of our uncertainty on $p_0$ might not be constant across the canonical parameter space; for instance we could assume more uncertainty in the tails of the distribution $p_0$ than in its body. 
We can interpret $\alpha(v)$ as a means to control uncertainty regarding the choice of the model $p_0$, and it is appropriately chosen based on the usage of the model. The level $\alpha(v)$ may also depend on the uncertainty surrounding the data and calibration.\\

Since $\mathcal{U}$ is a subset of the normal neighbourhood around $p_0$, the exponential map is well defined and we can construct a corresponding set of models close enough to $p_0$:
\begin{eqnarray*}
U:= \exp_{p_0}(\mathcal{U})= \{ p\in \mathcal{M}: d(p_0,p)\leq \alpha(v) \},
\end{eqnarray*}
From now on, we shall require the boundary $\partial U=\{ \alpha(v)\,v \; | \; v\in \mathcal{S}(p_0, 1) \}$ to be continuous and piecewise regular. Moreover, $U$ shall be a \emph{star--shaped set with respect to $p_0$} that is defined as follows:
\begin{definition}
A compact subset $U$ of a Riemannian manifold $\mathcal{M}$ is called star--shaped with respect to $p_0\in U$ if $\forall p\in U, p\neq p_0$ there exists a minimizing geodesic $\gamma$ with $\gamma(0)=p_0$ and $\gamma(T_p)=p$ such that $\gamma(t)\in U$ for all $t\in [0, T_p]$, where $T_p> 0$. 
\end{definition}

One advantage of the exponential map in this setting is that we can avoid calibration of different alternative models inside $U$. For each unit vector $v\in \mathcal{U}$ there exists a unique geodesic connecting points on the boundary of $U$ with the point $p_0$. This geodesic is given by $\gamma(t)=\exp_{p_{0}}(tv)$ for $t\in [0, \alpha(v)]$.

\subsection*{Example}
Demonstrating that the model is suitable for the intended purpose is a critical part of the analysis of model risk. 
We want to evaluate the impact of relaxing the assumption of symmetry for the underlying $P\& L$ distribution, i.e. the impact of not including the skew in the model. Hence, we embed the model manifold $\mathcal{M}$ into a larger manifold of skew--normal distributions, $\bar{\mathcal{M}}=\{ p(x; \mu, \sigma, s): \mu \in \mathbb{R}, \sigma>0, s\in \mathbb{R}\}$, where 
$s$ is the shape parameter (\cite{azzalini1985class}). Note that for $s=0$ we re--obtain the initial normal distribution, $\mathcal{N}(\mu, \sigma)$. The skew normal distribution family takes into account the skewness property. \\

After considering various time windows, data sequences, fittings and estimates, we determine the neighbourhood of our model to be the geodesic connecting the base model, $p_0=\mathcal{N}(2,10)=\mathcal{SN}(2,10,0)$, and the skew--normal distribution, $p_1=\mathcal{SN}(\mu, \sigma, s)$, with parameters $\mu=1.95, \sigma=9.98$, and $s=2$. The geodesic distance between these two distributions is $d(\sqrt{p_0}, \sqrt{p_1})=0.6809$. To form the neighbourhood, we first construct the related perturbation tangent vector associated with the directions to the boundary point, $\sqrt{p_1}$, using the inverse exponential map defined by \ref{eq:exponential}
\begin{eqnarray*}
	v_{p_1} &=& \exp^{-1}_{\sqrt{p_0}} \big( \sqrt{p_1} \big)
= \Big[ \dfrac{0.6809}{\sin(0.6809))}\Big(\sqrt{p_1}-\cos\big(0.6809)\big)\sqrt{p_0}\Big) \Big]
	\end{eqnarray*} 
This provides a class of variations of the initial model by moving away from it in the direction $v_{p_1}$ which determines the whole neighbourhood $U$ given by
\begin{eqnarray*}
	U=\{ \gamma(t)=(\exp_{\sqrt{p_0}}(t v_{p_1}))^2; t \in [0, 1] \}
\end{eqnarray*}
with boundaries $\partial_1 U=\{ p_0 \}$ and $\partial_2 U=\{ p_1 \}$
Thus, by varying $t$ from $0$ to $1$, one traces the geodesic path from $p_0$ to $p_1$, and we obtain a set of all distributions in the direction $p_1$. 
The neighbourhood $U$ around $p_0$ includes all distributions on the geodesic $\gamma$ for $t\in [0,1]$.\\

\QEDB

\section{Weight Function Definition}

Variations of the chosen model are not equally material and they all might take place with different probabilities. By placing a non--linear weight function (kernel), $K$, over the set $U$ we can easily place relative relevance to each alternative model, and assign the credibility of the underlying assumptions that would make alternative models partially or relatively preferable to the nominal one $p_0$. The particular choice of the structure of the kernel depends on various factors, such as usage of the model, distance from $p_0$, or sensitivity to different changes.\\ 

In what follows we define a general weight function $K$ and show that under certain conditions it is well defined and unique. In general, we consider $K$ to be a non--negative and continuous function that depends on the local geometry of $\mathcal{M}$ by incorporating a Riemannian volume associated to the Fisher--Rao information metric given by $dv(p)=\sqrt{det(I(\theta))}d\theta$. The volume measure is the unique Borel measure on $\mathcal{M}$ (\cite{federer2014geometric}). With respect to a coordinate system, the information density of $p$ represents the amount of information the single model possesses with respect to the parameters. For example, a small $dv(p)$ means that the model contains much uncertainty and requires many observations to learn. 
\\

As the underlying factors\footnote{For example the uncertainty surrounding data, calibration or model selection.} that influence the perturbations of the given model happen with some likelihood, we treat all models inside $\mathcal{M}$ as random objects. As a consequence, we require $K$ to be a probability density with respect to the Riemannian volume, i.e. $\int_{\mathcal{M}} Kdv(p)=1$. 
Additionally, we state that the right model does not exist and that the choice of $p_0$ was to some extent a subjective preference. 
\begin{definition}\label{def:kernel}
An admissible weight function $K$ defined on $\mathcal{M}$ satisfies the following properties:
\begin{description}
\item[($K1^{\prime}$)] $K$ is continuous on $\mathcal{M}$ 
\item[($K2^{\prime}$)] $K\geq 0$ for all $p\in \mathcal{M}$
\item[($K3$) ] $\int_{\mathcal{M}} Kdv(p)=1$
\end{description}
\end{definition}

Recall that to compute the $n$--dimensional volume of the objects in $\mathcal{M}$, one considers a metric tensor on the tangent space $T_{p}\mathcal{M}$ at $p\in \mathcal{M}$. In particular, the Fisher--Rao information metric $I$ on $\mathcal{M}$ maps each $p\in \mathcal{M}$ to a volume $dv(p)$ which is a symmetric and bilinear form that further defines an $n$--dimensional volume measure on any measurable subset $U\subset \mathcal{M}$ by $V\!ol(U):=\int_{U} dv(p)$. A smooth probability density $K$ over $\mathcal{M}$ with respect to the Riemannian measure induces a new absolutely continuous probability measure $\zeta$ with respect to $V\!ol$
\begin{eqnarray}\label{eq:measure}
\zeta(U)=\int_{U}d\zeta=\int_{U}Kdv(p)
\end{eqnarray}
for all measurable $U\subset \mathcal{M}$ and $\zeta(\mathcal{M})=1$.  The pair $(\mathcal{M}, \zeta)$ is then called a weighted manifold, or a Riemannian metric--measure space and is proved to be a nontrivial generalization of Riemannian manifolds (\cite{morgan2005manifolds}). 
\\

The weight function $K$ of the Definition \ref{def:kernel} represents a general characterization of a probability density over the Riemannian manifold $\mathcal{M}$. To tune $K$ for proper analysis of model risk, we need to impose additional properties which are connected with the specific uncertainties surrounding the given model. \\

From a practitioner point of view, models that do not belong to the chosen neighbourhood $U$ are not relevant from the perspective of model risk, and so do not add any uncertainty. Therefore, we assume the weight function to be non--negative only over the neighbourhood $U$ and zero elsewhere. Moreover, translation of the changes in various underlying assumptions, data or calibration into the changes in output and further usage of the model are going to vary with respect to the direction of the change. Hence, we require $K$ to be continuous along the geodesic curves $\gamma$ uniquely determined by $v\in \mathcal{S}(p_0,1)\subset T_{p_0}U$ starting at $p_0$ and ending at the points on $\partial U$. These additional properties are a modification of $(K1^{\prime})$ and $(K2^{\prime})$:
{\it 
\begin{description}
\item[($K1$)] $K$ is continuous along all geodesics $\gamma$ starting at $p_0$ for all unit vectors on $\mathcal{S}(p_0,1)$ 
\item[($K2$)] $K> 0 \,\, \forall p\in U\backslash \{ \partial U \}$ and $K\geq 0 \,\, \forall p\in \partial U$, and $K=0 \,\, \forall p\in \mathcal{M}\backslash \{ U\}$
\end{description}
}

The weight function satisfying properties $(K1)-(K3)$ takes into consideration and is adjusted according to the different directions of the changes, i.e. prescribes different sensitivities to different underlying factors.

\section{Weight Function Construction}
The construction of a weight function on a given Riemannian manifold is technically difficult since it requires precise knowledge of the intrinsic geometry and the structure of the manifold. To determine a weight function $K$ that satisfies all of the required properties and in order to overcome this difficulty we introduce a continuous mapping from a manifold endowed with an Euclidean geometry to the model manifold endowed with a Riemannian geometry that preserves the local properties. Euclidean geometry is well understood and intuitive, and thus a construction of a function on this space is considerably easier and more intuitive. In total, we construct three mappings:  the exponential map $\exp_{p_0}$, the polar transform $P$ and a further coordinate transform $\Lambda_{\rho}$.
\\

Every Riemannian manifold $\mathcal{M}$ is locally diffeomorphic to the Euclidean space $\mathbb{R}^n$, and so in a small neighbourhood of any point the geometry of $\mathcal{M}$ is approximately Euclidean. All inner product spaces of the same dimension are isometric, therefore, all the tangent spaces $T_{p}\mathcal{M}$ on a Riemannian manifold $\mathcal{M}$ are isometric to the $n$--dimensional Euclidean space $\mathbb{R}^n$ endowed with its canonical inner product. Hence, all Riemannian manifolds have the same infinitesimal structure not only as manifolds but also as Riemannian manifolds.\\

The weight function is defined with respect to the neighbourhood $U$ and is continuous on the geodesic curves $\gamma$ connecting $p_0$ to the points on the boundary $\partial U$. All material perturbations, i.e. alternative models inside $U$, are uniquely described by the distances from $p_0$ and by the vectors tangent to the unique geodesics $\gamma$ that pass through them. To maintain these properties, we consider an $n$--dimensional cylinder $C^n=[0,1]\times \mathbb{S}^{n-1}=\{ (t,\nu): t\in [0,1], \nu \in \mathbb{S}^{n-1} \}\subset \mathbb{R}^{n+1}$, where the parameter $t$ stands for the normalized distance of geodesics, and where $\mathbb{S}^{n-1}$ denotes the $(n-1)$--dimensional unit sphere on $\mathbb{R}^n$ containing all the unit tangent vectors of $T_{p_0}\mathcal{M}$. The boundaries of $C^n$ are
\begin{eqnarray*}
\partial_1 C^n =\{ (0,\nu) :  \nu \in \mathbb{S}^{n-1} \}, \,\,\,\,\, \partial_2 C^n =\{ (1,\nu) :  \nu \in \mathbb{S}^{n-1} \},
\end{eqnarray*}
and represent the end points of the geodesics, i.e. $\partial_1 C^n$ will be transformed into $p_0$ and $\partial_2 C^n$ into $\partial U$.\\
 
The Riemannian structure on $C^n$ is given by the restriction of the Euclidean metric in $\mathbb{R}^{n+1}$ to $C^n$. Hence, $C^n$ is a compact smooth Riemannian manifold with a canonical measure given by the product measure $dt\times d\nu$. This manifold allows us to construct an appropriate function on $C^n$, and then obtain a weight function satisfying all required properties $(K1)-(K3)$. 
\\ 

As a first step to obtain a mapping from $C^n$ to $\mathcal{M}$, we consider the exponential map from the tangent space at the point $p_0$ onto the neighbourhood $U$. Since $U$ is compact and, hence, topologically complete, the geodesic $\gamma$ can be defined on the whole real line $\mathbb{R}$ (\cite{HopfRinow}). Thus, the exponential map is well--defined on the whole tangent space $T_{p_0}\mathcal{M}$. Further, since $U$ is a subset of the normal neighbourhood of $p_0$, the exponential map defines a local diffeomorphism from $T_{p_0}U$ to $U$. Then the geodesics $\gamma$ are given in these coordinates by rays emanating from the origin.

\subsection*{Example}
The weight function is constructed with respect to the neighbourhood $U$ that in our example represents the geodesic $\gamma$ with boundary points $p_0$ and $p_1$. We parametrize $\gamma$ by $t\in [0,1]$, and define the one--dimensional cylinder as $ C^1= [0,1]\times \mathbb{S}$ with boundaries $\partial_1 C^1=\{ (0, \nu) \}$ and $\partial_2 C^1=\{ (1, \nu) \}$ where $\mathbb{S}=\{\nu\in \mathbb{R}^n:  ||\nu||^2=\nu^2=1\}$. The left boundary $\partial_1 C^1$ will be contracted to the given model $p_0$ and $\partial_2 C^1$ to $p_1$. 

The construction of the weight function reduces to the construction of the weight function on the real line on the interval $[0, 1]$. 

\QEDB
\\

Next, we introduce a polar coordinate transformation on $T_{p_0}\mathcal{M}$. For the sake of distinction, we denote by $\mathcal{S}(p_0, 1)$ the $(n-1)$--dimensional unit sphere in $T_{p_0}\mathcal{M}$, and by $\mathbb{S}^{n-1}$ the unit sphere in $\mathbb{R}^n$:
\begin{align}\label{eq:polar}
\begin{split}
& P:\,\, [0, \infty)\times \mathcal{S}(p_0, 1) \rightarrow T_{p_{0}}\mathcal{M}: \,\,\, P(t, v)=tv; \\
&  P^{-1}:\,\, T_{p_{0}}\mathcal{M}\backslash \{ 0\} \rightarrow (0, \infty)\times \mathcal{S}(p_0, 1) : \,\,\, P^{-1}(v)=\Big(||v||, \dfrac{v}{||v||}\Big)
\end{split}
\end{align}

In order to precisely describe the neighbourhood $U$, we define $\rho(v)$  as the length $d(p_0,p)\geq 0$ of the geodesic $\gamma$ connecting $p_0$ with the boundary $p\in \partial U$ in the direction $v\in \mathcal{S}(p_0, 1)$. The distance $\rho$, considered as a real valued function on $\mathcal{S}(p_0,1)$, is strictly positive and Lipschitz continuous on $\mathcal{S}(p_0,1)$. We now define the coordinate transformation
\begin{eqnarray*}
\Lambda_{\rho} :  (t, v)\mapsto \Big( \rho(v)t, v \Big).
\end{eqnarray*}
The mapping $\mathbb{S}^{n-1} \to \mathcal{S}(p_0,1)$, $\nu \mapsto v$ is well defined in the sense that there exists a canonical identification between a unit vector $\nu\in \mathbb{S}^{n-1}\subset \mathbb{R}^n$ and the element $v\in \mathcal{S}(p_0,1)\subset T_{p_0}\mathcal{M}$. Since the distance $v\mapsto\rho(v)$ is strictly positive and Lipschitz continuous on $\mathcal{S}(p_0, 1)$, so is the inverse $v\mapsto \dfrac{1}{\rho(v)}$. Therefore, the mapping $\Lambda_{\rho}$ defines a bi--Lipschitz mapping from $[0,1]\times \mathbb{S}^{n-1}$ onto the subset $[0, \rho(v)]\times \mathcal{S}(p_0, 1)$.\\

Therefore the composition $\exp_{p_0}P \Lambda_{\rho}$ defines a mapping from $C^n$ onto $U$ that maps $\partial_1 C^n$ onto the point $\{ p_0\}$ and the right hand side boundary onto $\partial U$. Moreover, it preserves continuity for any continuous function $h$ defined on $C^n$ that satisfies the following consistency condition: 
\begin{definition}\label{def:consistency}
A continuous function $h$ defined on a cylinder $C^n$ is called consistent with a continuous function $f$ on $U$ under the mapping $\exp_{p_{0}}P\Lambda_{\rho}$ if $h(t,\nu)=\Lambda_{\rho}^{-1}f^{-1}(t,\nu)$ for all $(t,\nu)\in C^n$. In this case, $h$ satisfies the following conditions:
\begin{eqnarray*}
(i) \,\, h(0,\nu_1)&=& h(0,\nu_2) \,\,\, \forall \nu_1, \nu_2\in \mathbb{S}^{n-1}\\
(ii) \,\, h(1,\nu_1)&=& h(1,\nu_2) \,\,\, \text{if} \,\,\,\exp_{p_0}P\Lambda_{\rho}(1,\nu_1)=\exp_{p_0}P\Lambda_{\rho}(1,\nu_2)\,\, \text{on}\,\, \mathcal{M}
\end{eqnarray*}
\end{definition} 
The first condition $(i)$ implies that $h$ is constant on the boundary $\partial_1 C^n$. When the function $h$ on $C^n$ is consistent with $f$, the constant value at $\partial_1 C^n$ corresponds exactly with the value $f(p_0)$. The second condition ensures compatibility of function $h$ with function $f$ at the points of the boundary $\partial U$, i.e. if $\exp_{p}P\Lambda_{\rho}$ maps two different points $(1, \nu_1)$ and $(1, \nu_2)$ in $C^n$ onto the same point $p\in \partial U$, then $h(1, \nu_1)=h(1, \nu_2)=f(p)$. 

\begin{lemma}
The existence of the weight function $K$ satisfying assumptions $(K1)-(K3)$ is equivalent to assuming the existence of a consistent function $h(t,\nu)$ defined on $C^n$ with codomain $\mathbb{R}^n$ satisfies the following properties:
\begin{description}
\item[(H1)] $h(t,\nu)$ is a continuous function on the compact manifold $C^n$
\item[(H2)] $h(t,\nu)\geq 0, \,\, (t,\nu)\in [0,1)\times \mathbb{S}^{n-1}$ 
\item[(H3)] $h(1,\nu)=\kappa(\nu)$\, for all $\nu\in \mathbb{S}^{n-1}$, where $\kappa$ is some non--negative function of $\nu$
\item[(H4)] $h(0,\nu_1)=h(0,\nu_2)=const.$\, for all $\nu_1, \nu_2\in \mathbb{S}^{n-1}$
\item[(H5)]  $\int_{C^n}h(t,\nu)d\nu=1$, where $d\nu=dt\times d\mu$
\end{description}
\end{lemma}
See the Appendix for a proof. Using this result, the construction of the weight function becomes easier and more intuitive. One chooses the appropriate function $h$ defined on $C^n$ with respect to the particular model and the uncertainty surrounding it. Then, applying the above transformation one obtains an appropriate weight function $K$ defined on $U$ satisfying properties $(K1)-(K3)$ relevant for model risk analysis. Besides, for a chosen function $h$ the weight function $K$ is unique and well defined. 
\begin{theorem}
A continuous function $h$ defined on $C^n$ satisfying conditions $(H1)-(H5)$, determines a unique and well defined weight function $K$ satisfying $(K1)-(K3)$ on $U$ given by
\begin{eqnarray}\label{eq:kernel}
K(p,t)=\dfrac{1}{\eta_{p_0}(p)}t^{1-n}\rho(v)^{-n}h\Bigg(\dfrac{t}{\rho(v)},v \Bigg)
\end{eqnarray}
where $\eta_{p_0}(p)$ is the volume density with respect to $p_0$, $v$ is the tangent vector, $t\in [0,\rho(v)]$ is a scaling parameter, and $\rho(v)$ is the distance function defined above. 
\end{theorem}

\subsection*{Example}
In line with our example, we construct a suitable weight function adjusted to the uncertainty surrounding the VaR model. We have seen in the previous section that the underlying process suggests small deviations from the normal distribution and indicate a negative skew. Thus, to determine the weight function we construct a continuous function $h$ that has the maximum value at the point representing our given model $p_0$, and is monotonically--decreasing with the distance from $p_0$. This choice means that we are interested more on how sensitive is the model to small variations around $p_0$. We define $h$ on $[0,1]$ as follows
\begin{eqnarray*}
h(t)= c\Big(1-t\Big), \,\,\,\,\,\,  t\in [0,1]
\end{eqnarray*}
where the normalizing constant $c$ ensures the assumptions $(H5)$ and equals\footnote{ The volume of the $(n-1)$--dimensional ball $S(0,1)$ is $ 2 \pi^{1/2} \backslash \Gamma\bigg(\dfrac{1}{2}\bigg)$. Thus we have
\begin{eqnarray*}
1=  c \int_{[0,1]\times S^{n-1}} \Big(1-t\Big)dt\times d\mu\,\,\,\,\,\, \Rightarrow \,\,\,\,\,\, c= \Gamma\Big( \dfrac{n}{2} \Big) \pi^{-\pi/2}
\end{eqnarray*}} to $\Gamma\bigg(\dfrac{n}{2}\bigg)\pi^{-n/2}$. Note that since we have only one tangent vector $\nu$, $h$ depends only on the parameter $t$. By applying the continuous mapping $\exp_{p_0}P\Lambda_{\rho}$ we obtain the weight function $K$ along the geodesics $\gamma$:
\begin{eqnarray*}
K(p, t)&=& \dfrac{1}{\eta_{p_0}(p)}d(p_0, p_1)^{-1} \Gamma\bigg(\dfrac{1}{2}\bigg)\pi^{-1/2} \Bigg(1-\dfrac{t}{d(p_0, p_1))} \Bigg)\\
&=& 1.47 \Big(1-1.47t \Big)
\end{eqnarray*}
\QEDB


\section{Measure of Model Risk}

In this section we shall introduce a mathematical definition of the quantification of model risk, relate it to the concepts introduced so far and study some actual applications.\\

Recall that we have so far focused on a weighted Riemannian manifold $(\mathcal{M}, I, \zeta)$ with $I$ the Fisher--Rao metric and $\zeta$ as in eq.~\ref{eq:measure}. The model in previous sections was assumed to be some distribution $p \in \mathcal{M}$. More likely, a practitioner would define the model as some mapping $f: \mathcal{M} \to \mathbb{R}$ with $p \mapsto f(p)$, i.e. a model outputs some quantity.\footnote{This is not always the case but we can proceed along these lines depending on the usage to be given to the quantification itself. For example, an inter(extra)polation methodology on a volatility surface is a model whose output is another volatility surface, not a number. If we want to quantify the model risk of that particular approach for Bermudans we might consider its impact on their pricing. }\\

We shall formally introduce the normed space $(\mathcal{F}, \left\Vert \cdot \right\Vert)$ such that $f \in \mathcal{F}$. Though not strictly necessary at this informal stage we shall assume completeness so $(\mathcal{F}, \left\Vert \cdot \right\Vert)$ is a Banach space.

\begin{definition}
\label{def:modelRisk}
With notation as above, let $(\mathcal{F}, \left\Vert \cdot \right\Vert)$ be a Banach space of measurable functions with respect to $\zeta$. The \emph{model risk} $Z$ of $f \in \mathcal{F}$ and $p_0$ is given by
\begin{equation}
  \label{eq:7_1}
  Z(f,p_0) = \left\Vert f-f(p_0) \right\Vert.
\end{equation}
\end{definition}
Note that the measure represents the standard distance. All outcomes are constrained by the assumptions used in the model itself and so, the model risk is related to the changes in the output while relaxing them. The relevant model risk is therefore the difference between two models, rather than a hypothetical difference between a model and the truth of the matter.\\

The quantification of model risk itself can be thought of as a model with a purpose such as provisions calculation or comparison of modelling approaches. Possibilities are endless so we might have started with some $T:\mathcal{F} \to \mathcal{F}$ and set $Z(f,p_0) = \left\Vert T \circ f \right\Vert$\footnote{For example, another possibility is to use $\left\Vert \frac{f )}{f(p_0)} \right\Vert$ or $\left\Vert \frac{f - f(p_0)}{f(p_0)} \right\Vert$. These functional forms would allow us to obtain a dimensionless number which is might be a desirable property.}; however, we think eq.~\ref{eq:7_1} is general enough for our present purposes.\\

In what follows we address four examples of Def.~\ref{def:modelRisk}. Their suitability very much depends among other factors on the purpose of the quantification, as we shall see below.

\begin{enumerate}
\item $Z^1(f, p_{0})$ for $f\in L^1(\mathcal{M})$ represents the total relative change in the outputs across all relevant models:
$$Z^1(f, p_{0}) = \left\Vert f-f(p_0) \right\Vert_{1} =  \int_{\mathcal{M}} \Big|f-f(p_0) \Big|d\zeta$$

\item $Z^2(f, p_{0})$ for $f\in L^2(\mathcal{M})$ puts more importance on big changes in the outputs (big gets bigger and small smaller). It would allow to keep consistency with some calibration processes such as the maximum likelihood or least square algorithms:
$$Z^2(f, p_{0}) = \left\Vert f-f(p_0) \right\Vert_{2} = \Big(\int_{\mathcal{M}} \Big(f-f(p_0) \Big)^2 d\zeta \Big)^{1/2}$$

\item $Z^{\infty}(f, p_{0})$ for $f\in L_{\infty}(\mathcal{M})$ finds the relative worst--case error with respecto to $p_0$:
$$Z^{\infty}(f, p_{0}) =  \left\Vert f-f(p_0) \right\Vert_{\infty} = \esssup_{\mathcal{M}} \Big| f-f(p_0) \Big|$$
Further, it can point to the sources of the largest deviances: Using $\exp^{-1}_{p_0}$ we can detect the corresponding direction and size of the change in the underlying assumptions.
\item $Z^{s,p}(f, p_{0})$ for $f\in W^{s,p}(\mathcal{M})$ is a Sobolev norm that can be of interest in those cases when not only $f$ is relevant but its rate of change:\footnote{An example can be a derivatives model used not only for pricing but also for hedging.}
$$Z^{s,p}(f, p_{0})= \left\Vert f-f(p_0) \right\Vert_{s,p}=\Bigg(\sum_{|k|\leq s} \int_{\mathcal{M}} \Big|\partial^{k}\Big(f-f(p_0)\Big) \Big|^p d\zeta \Bigg)^{1/p}$$
\end{enumerate}

Sound methodology for model risk quantification should at least consider the data used for building the model, the model foundation, the IT infrastructure, overall performance, model sensitivity, scenario analysis and, most importantly, usage. Within our framework we address and measure the uncertainty associated with the aforementioned areas and the information contained in the models. The choice of the embedding and proper neighbourhood of the given model take into account the knowledge and the uncertainty of the underlying assumptions, the data and the model foundation. The weight function that assigns relative relevance to the different models inside the neighbourhood considers the model sensitivity, scenario analysis, the importance of the outcomes with connection to decision making, the business, the intended purpose, and it addresses the uncertainty surrounding the model foundation. Besides, every particular choice of the norm provides different information of the model. Last and most important, the model risk measure considers the usage of the model represented by the mapping $f$.\footnote{Or equivalently by any possible transformation $T:\mathcal{F}\rightarrow \mathcal{F}$.}

\section{Conclusions and Further Research}

In this paper we introduce a general framework for the quantification of model risk using differential geometry and information theory. We also rigorous a sound mathematical definition of model risk using Banach spaces over weighted Riemannian manifolds, applicable to most modelling techniques using statistics as a starting point.\\

Our proposed mathematical definition is to some extent comprehensive in two complementary ways. First, it is capable of coping with relevant aspects of model risk management such as model usage, performance, mathematical foundations, model calibration or data. Second, it has the potential to asses many of the mathematical approaches currently used in financial institutions: Credit risk, market risk, derivatives pricing and hedging, operational risk or XVA (valuation adjustments).\\

It is worth noticing that the approaches in the literature, to our very best knowledge, are specific in these same two ways: They consider very particular mathematical techniques and are usually very focused on selected aspects of model risk management.\\

There are many directions for further research, all of which we find to be both of theoretical and of practical interest. We shall finish by naming just a few of them:

Banach spaces are very well known and have been deeply studied in the realms of for example functional analysis. On the other hand, weighted Riemannian manifolds are non--trivial extensions of Riemannian manifolds, one of the building blocks of differential geometry. The study of Banach spaces over weighted Riemannian manifolds shall broaden our understanding of the properties of these spaces as well as their application to the quantification of model risk.

Our framework can include data uncertainties by studying perturbations and metrics defined on the sample, which are then transmitted to the weighted Riemannian manifold through the calibration process.

The general methodology can be tailored and made more efficient for specific risks and methodologies. For example, one may interpret the local volatility model for derivatives pricing as an implicit definition of certain family of distributions, extending the Black--Scholes stochastic differential equation (which would be a means to define the lognormal family).

Related to the previous paragraph, and despite the fact that there is literature on the topic, the calculation of the Fisher--Rao metric itself deserves further numerical research in order to derive more efficient algorithms.

\section{Appendix}

In the Appendix we present the proof of Lemma 1 and Theorem 4.  \\

{\it Proof of Lemma 1.}
To prove the equivalence we need to show that the function $h$ defined in Lemma \ref{lemma:L1} preserves the required properties of $K$ under the continuous mapping $\exp_{p_0}P\Lambda_{\rho}$. First we show that the composition that consists of three different mappings is well defined. \\

As a first step, we define an $n$--dimensional cylinder
\begin{eqnarray*}
C^n:=[0, 1]\times \mathbb{S}^{n-1}=\{ (t,\nu): t\in [0, 1], \nu\in \mathbb{S}^{n-1} \}\subset \mathbb{R}^{n+1}
\end{eqnarray*} 
where $\mathbb{S}^{n-1}:=\{ \nu\in \mathbb{R}^n: ||\nu||^2=\nu_1^2+\dots+\nu_n^2=1 \}$ denotes the $(n-1)-$dimensional unit sphere in $\mathbb{R}^n$. The cylinder $C^n$ is a differentiable submanifold of $\mathbb{R}^{n+1}$ with boundaries
\begin{eqnarray*}
\partial_{1}C^n :=\{ (0, \nu): \nu\in \mathbb{S}^{n-1} \}, \,\,\, \partial_{2}C^n :=\{ (1, \nu): \nu\in \mathbb{S}^{n-1} \}
\end{eqnarray*} 
A Riemannian structure on $C^n$ is given by the restriction of the Euclidean metric in $\mathbb{R}^{n+1}$ to $C^n$. Thus, $C^n$ is a compact Riemannian manifold. A canonical measure on $C^n$ is given by the product measure $dt\times d\mu(\nu)$, where $\mu$ denotes the standard surface measure on $\mathbb{S}^{n-1}$.
\\

We define $\rho(v)$ as the length $d(p_0,p)\geq 0$ of the geodesic $\gamma$ connecting the point $p_0$ with a boundary point $p\in \partial U$ in the direction $v\in \mathcal{S}(p_0, 1)$, where $\mathcal{S}(p_0, 1)$ denotes the $(n-1)-$dimensional unit sphere in the tangent space $T_{p_{0}}\mathcal{M}$.\\

Note that since $U$ is the subset of the normal neighbourhood with respect to $p_0$, the exponential map is isometric.  From now on we will assume that $U$ is a compact star--shaped subset of a Riemannian manifold $\mathcal{M}$ and the distance function $\rho$ is Lipschitz continuous on $S(p_0, 1)\subset T_{p_0}\mathcal{M}$. Lipschitz continuity of $\rho(v)$ is equivalent to the assumption of continuity and piecewise regularity of $\partial U$. \\

Now we define an $n-$dimensional subset of $C^n$ by
\begin{eqnarray*}
C_{\rho}^n:= \{ (t,v): t\in [0, \rho(v)], v\in \mathbb{S}^{n-1} \}\subset [0, 1]\times \mathbb{S}^{n-1}
\end{eqnarray*} 
with boundary
\begin{eqnarray*}
\partial_{1}C_{\rho}^n :=\{ (0, v): v\in \mathbb{S}^{n-1} \},\,\,\, \partial_{2}C_{\rho}^n :=\{ (\rho(v), v): v\in \mathbb{S}^{n-1} \}
\end{eqnarray*}
The new set $C_{\rho}^n$ is a compact subset of $C^n$. In order to map $C^n$ onto $C_{\rho}^n$ we define the following coordinate transform:
\begin{eqnarray*}
\Lambda_{\rho} : C^n \rightarrow C_{\rho}^n, \,\,\,\, (t, v)\rightarrow \Big( \rho(v)t, v \Big).
\end{eqnarray*}
Since the distance function $v\mapsto\rho(v)$ is strictly positive and Lipschitz continuous on $\mathcal{S}(p_0, 1)$, so it is the inverse function $v\mapsto \dfrac{1}{\rho(v)}$. Therefore, the mapping $\Lambda_{\rho}$ defines a bi--Lipschitz mapping from $C^n$ onto $C_{\rho}^n$.
The Jacobian determinant of $\Lambda_{\rho}$ equals $\rho$ almost everywhere on $C^n$. 
\\

Next we consider the polar transformation $P$ defined by equation \ref{eq:polar} which is well defined by continuity in $T_{p_0}\mathcal{M}$, and maps $C_{\rho}^n$ onto $\mathcal{U}\subset T_{p_0}\mathcal{M}$. Moreover, the transformation $P$ defines a diffeomorphism from $C_{\rho}^n\backslash \{\partial_1 C_{\rho}^n, \partial_2 C_{\rho}^n\}$ onto the open set $\mathcal{U}\backslash \{0, \partial \mathcal{U}\}$. Combining $P$ with the exponential map $\exp_{p_0}$, we have
\begin{eqnarray*}
\exp_{p_0}P(C_{\rho}^n)=U
\end{eqnarray*}
The composition $\exp_{p_0} \circ P$ defines a diffeomorphism from $C_{\rho}^n\backslash \{ \partial_1 C_{\rho}^n, \partial_2 C_{\rho}^n  \}$ onto $U\backslash \{ p_0, \partial U \}$. Furthermore, the boundary $\partial_1 C_{\rho}^n$ is mapped onto $\{p_0\}$ and the boundary $\partial_2 C_{\rho}^n$ onto the boundary $\partial U$~\footnote{When $p_0\in \partial U$, the boundary  $\partial_2 C_{\rho}^n$ is mapped onto $\partial U\backslash \{p_0\}$}. Then the points $(t,v)\in C_{\rho}^n$ induce geodesic polar coordinates on $\mathbb{R}^n$. 
\\

We have introduced three mappings
\begin{eqnarray*}
C^n \xrightarrow{\Lambda_{\rho}} C_{\rho}^n \xrightarrow{P} \mathcal{U} \xrightarrow{\exp_{p_{0}}} U \subset \mathcal{M}
\end{eqnarray*}
The composition $\exp_{p_0}P\Lambda_{\rho}$ is a continuous mapping from $C^n$ onto $U$. Moreover, $\exp_{p_0}P\Lambda_{\rho}$ maps the boundary $\partial_1 C^n$ of the cylinder $C^n$ onto the point $p_0$ and the boundary $\partial_2 C^n$ onto $\partial U$. 

Now we prove that a consistent function satisfying properties $(H1)-(H5)$ uniquely determines the weight function satisfying $(K1)-(K3)$: 
\begin{itemize}
\item It is straightforward to see, that properties $(K1)-(K2)$ are satisfied by construction. The composition $\exp_{p_0}P\Lambda_{\rho}$ preserves connectedness and compactness and it a continuous mapping from $C^n$ onto $\mathcal{M}$. Moreover, $\exp_{p_0}P\Lambda_{\rho}$ maps the left hand boundary $\partial_1 C^n$ onto the point $p_0$ and the right hand boundary $\partial_2 C^n$ onto the boundary of $U$. Hence, the image $f(\exp_{p_0}(\rho(v)tv))$ of a continuous function $f$ on $\mathcal{M}$ is also continuous on the cylinder $C^n$, and every function $g$ defined on $C^n$ satisfying consistency properties $(i)-(ii)$ of Def.\ref{def:consistency}  is the image of a continuous function on $\mathcal{M}$ under the pull--back operator $\Lambda^{-1}_{\rho}P^{-1}\exp_{p_0}^{-1}$.   

The composition $\exp_{p_0}P\Lambda_{\rho}$ applied to a function $h$ that is continuous on $C^n$ and satisfies the consistency conditions $(i)-(ii)$ of Def.\ref{def:consistency}  will give us a continuous function $K$ on $\mathcal{M}$ that by construction is continuous along the geodesics starting at $p_0$ and ending at the points of the boundary. That means, property $(K1)$ is satisfied. The same argument applies to any non--negative function $h$ on $C^n$. Thus, properties $(H1)-(H2)$ ensures $(K1)-(K2)$ under the composition $\exp_{p0}P\Lambda_{\rho(v)}$.
\item Further, it remains to prove that the weight function $K$ is indeed a probability density on $\mathcal{M}$ with respect to the measure $dv(p)$, i.e. to show that $\int_{M} d\zeta=1$.
\begin{eqnarray*}
\int_{\mathcal{M}} d\zeta=\int_{\mathcal{M}} K(p,t)dv(p)=\int_{T_{p_0}\mathcal{M}} K(\exp_{p_0}(v), t)\eta(v)d\xi
\end{eqnarray*}
where $d\xi$ is the standard Lebesgue measure on the Euclidean space $T_{p_0}\mathcal{M}$ and $\eta_{p_0}(v)=det((d\exp_{p_0})_v)$ is the Jacobian determinant of the exponential map. Note that $\eta(v)$ represents the density function that is a positive and continuously differentiable function on $U\subset T_{p_0}\mathcal{M}$ and the zeros of $\eta$ lie at the boundary of $\mathcal{M}$. Further, we have
\begin{eqnarray*}
\int_{T_{p_0}\mathcal{M}} K(\exp_{p_0}(v), t)\eta(v)d\xi = \int_{\mathcal{S}(p_0, 1)}\int_{0}^{\rho(v)} t^{n-1} K(\exp_{p_0}(tv), t)\eta_{p_0}(tv)dt d\mu(v)
\end{eqnarray*}
where $t^{n-1}$ is the Jacobian determinant of the polar coordinate transformation and $d\mu(v)$ is the standard Riemannian measure on the unit sphere $\mathcal{S}(p_0, 1)$. The last step is the mapping from $C_{\rho}$ to $C^n$:
\begin{eqnarray*}
\int_{\mathcal{S}(p_0, 1)}\int_{0}^{\rho(v)} t^{n-1} K(\exp_{p_0}(tv))\eta_{p_0}(tv)dt d\mu(v) = \int_{S^{n-1}}\int_{0}^{1}\dfrac{1}{\rho(v)} K\Big( \exp_{p_0}(\rho(v) t v) \Big)\big(\rho(v)t\big)^{n-1}\eta_{p_0}(\rho(v)t v)dt d\mu(v)
\end{eqnarray*}
where the Jacobian determinant is $\dfrac{1}{\rho(v)}$. Then using the expression for $K$ we have that the expression above is equal to:
\begin{eqnarray*}
 && \int_{S^{n-1}}\int_{0}^{1}\dfrac{1}{\rho(v)} \dfrac{1}{\eta_{p_0}(\rho(v)t v)}t^{1-n}\rho(v)^{-n}h\Bigg(\dfrac{t}{\rho(v)},v \Bigg)\big(\rho(v)t\big)^{n-1}\eta_{p_0}(\rho(v)t v)dt d\mu(v)\\
&=& \int_{S^{n-1}}\int_{0}^{1} h\Bigg(\dfrac{t}{\rho(v)},v \Bigg)dt d\mu(v) = 1
\end{eqnarray*}
\end{itemize}
\QEDB

{\it Proof of Theorem 4.}
Note that the composition $\exp_{p_0}P\Lambda_{\rho}$ induces a change of variables for integrable function $f$ that yields to the following formula:
\begin{eqnarray*}
\int_{\mathcal{M}} f(p)d\zeta &=& \int_{U}  f(p)d\zeta =  \int_{\mathcal{U}}  f(\exp_{p_0}(v)) \eta_{p_0}(v)dv \\
 &=&  \int_{\mathcal{S}(p_0, 1)}\int_{0}^{\rho(v)}  f(\exp_{p_0}(tv)) t^{1-n}\eta_{p_0}(t,v)dtdv \\
 &=&   \int_{S^{n-1}}\int_{0}^{1} f(\exp_{p_0}(\rho(\nu)\nu)) \dfrac{1}{\rho(\nu)}\eta_{p_0}(t\rho(\nu),\nu)dtd\nu
\end{eqnarray*}

The volume density $\eta_{p_0}$ is a well--defined, non--negative function with zeros at the cut locus of the point $p_0$. Besides, $\eta_{p_0}$ is continuous and differentiable function on $\mathcal{M}$\footnote{Note that when $\mathcal{M}$ is $\mathbb{R}^n$ with the canonical metric, then $\eta_{p_0}(p)=1$ for all $p\in \mathbb{R}^n$.}. The distance function $\rho$ is a well defined, strictly positive and Lipschitz continuous function on $\mathcal{S}(p_0,1)$, and thus is the inverse $1/\rho(v)$. Therefore, the mapping $\Lambda_{\rho}$ defines a bi--Lipschitzian mapping from $C^n$ to $C^n_{\rho}$. Moreover, the composition $\exp_{p_0}P$ defines a diffeomorphism from $C^n_{\rho}\ \{ \partial_1 C^n_{\rho}, \partial_2 C^n_{\rho} \}$. The using the fact that the point set $\{p_0\}$ and the boundary of $U$ are subsets of $t\nu-$measure zero, we can conclude that the mapping $\exp_{p_0} P \Lambda_{\rho}$ is isomorphism. Then for any $h$ defined on $C^n$ satisfying conditions $(i)-(ii)$ of Def.\ref{def:consistency}, the associated weight function $K$ is well defined on $U$. The uniqueness of $K$ follows after specifying a function $h$ that satisfies properties $(H1)-(H5)$.  \QEDB

\newpage
\section*{References}

\end{document}